\documentclass[english]{article}
\usepackage[T1]{fontenc}
\usepackage[latin9]{inputenc}
\usepackage{array}
\usepackage{verbatim}
\usepackage{longtable}
\usepackage{textcomp}
\usepackage{graphicx}
\usepackage[authoryear]{natbib}

\makeatletter

\providecommand{\tabularnewline}{\\}

\newcommand{\lyxaddress}[1]{
\par {\raggedright #1
\vspace{1.4em}
\noindent\par}
}

\makeatother

\usepackage{babel}

\begin{document}

\title{On the Allometric Scaling of Resource Intake Under Limiting Conditions.}

\author{Alberto Basset$^{1,2}$\\
Francesco Paparella$^{2,3}$\\
Francesco Cozzoli$^{1}$}

\maketitle

\lyxaddress{$^{1}$DISTEBA - Università del Salento\\
$^{2}$Centro euroMediterraneo per i Cambiamenti Climatici\\
$^{3}$Dip. di Matematica {}``E. De Giorgi'' - Università del
Salento}


\pagebreak{}
\begin{abstract}
Individual resource intake rates are known to depend on both individual
body size and resource availability. Here, we have developed a model
to integrate these two drivers, accounting explicitly for the scaling
of perceived resource availability with individual body size. The
model merges a Kleiber-like scaling law with Holling functional responses
into a single mathematical framework, involving both body-size the
density of resources.

When the availability of resources is held constant the model predicts
a relationship between resource intake rates and body sizes whose
log-log graph is a concave curve. The significant deviation from a
power law accounts for the body size dependency of resource limitations.
The model results are consistent with data from both a laboratory
experiment on benthic macro-invertebrates and the available literature. 

\pagebreak{}
\end{abstract}

\section{Introduction}

Resource intake is a major component of individual fitness. This subject
was independently addressed from the perspectives of ecological energetics
\citep{Kleiber32}, niche theory \citep{Hutchinson59} and behavioral
ecology (e.g. the Charnov marginal value theorem; \citep{Charnov76}). 

It is well known that the rate at which individuals acquire resources
depends both on their body masses and on the overall abundance of
those resources that they can efficiently exploit. Ecological energetics
and, more recently, metabolic theory, have addressed the relationship
between resource intake rates and individual body mass \citep{Peters83,Brown04}.
Behavioral ecology has quantitatively addressed the relationships
between resource intake rate and resource availability, through the
so called {}``Holling's functional responses'' \citep{Holling59a,Holling59b}. 

In a resource limited environment, individual body size also affects
the individual perception of resource availability \citep{Haskell02}
determining patch selection \citep{Belowsky97,Ritchie98} and patch
departure behaviors \citep{Wilson99,Basset07}. The combined influence
of body size on individual metabolic rates and individual perception
of resource availability has been modeled as a major determinant of
interspecific coexistence \citep{Basset07}. 

A simple but far-reaching observation wafts in the literature: the
ingestion rate of large individuals is limited at higher resource
availability than that of smaller competitors: the former give up
the patch earlier and at higher densities of remaining resources than
the latter \citep{Brown94}, and they are more common in ecosystems
which are richer in nutrients and more productive \citep{Makarieva04};
consistently the body size of the largest species occurring in an
ecosystem has been found to be a growing function of the ecosystem
surface area \citep{Marquet05}. 

A body size dependency of resource availability has not yet been incorporated
in an allometric model relating metabolism to body mass. However,
when limiting resource conditions occur, it seems quite obvious that
larger individuals, or larger species, are limited earlier than their
smaller competitors, being comparatively less able to maintain optimal
resource acquisition rates. As a result, the actual resource acquisition
rates vs body-mass relationship should deviate from an ideal scaling
law, resulting in a concave curve in a log-log graph. 

Here, we propose a mathematical model that links the resource intake
rate with both body size and the level of available resources. The
model merges into a single framework a Kleiber-like scaling law, Holling's
functional response, and a scaling law linking the perceived level
of available resources with body size. Data from a laboratory experiment
on benthic macro-invertebrates and metadata from the available literature
are used to evaluate the realism of the model results.

\section{The Model \label{sec:The-Model}}

Kleiber's equation is a scaling law linking the metabolism to the
body size of individual living organisms. It can be written in the
following form:\begin{equation}
\frac{dQ}{dt}=P_{0}\left(\frac{M}{M_{0}}\right)^{\alpha}\label{eq:kleiber}\end{equation}
where:
\begin{description}
\item [{$Q$}] is the total energy required by the organism for its metabolic
needs;
\item [{$P_{0}$}] is a baseline power which determines the elevation of
the scaling law; 
\item [{$M$}] is the mass of the organism (or 'body size');
\item [{$M_{0}$}] is a baseline mass (so that the power law is applied
to a non-dimensional quantity);
\item [{$\alpha$}] is a positive constant, often taken to be equal to
$3/4$. 
\end{description}
There is a very strong consensus about the idea that $M$ and $dQ/dt$
are functionally related, but a considerable debate on the nature
of this relationship. A few theoretical models are consistent in deriving
the power-law in equation (\ref{eq:kleiber}) but they have spurred
much controversy about the underlying mechanisms \citep{West97,Banavar99,Makarieva04,Glazier05}.
The debate has often focused on the exact value to be given to the
scaling exponent, and the value $\alpha=3/4$ is the most commonly
cited one (e.g. \citealp{Peters83}). Different values have been observed
for different organisms or indicators of metabolism; see, e.g., \citep{Reich06,Enquist07}
for plants, \citep{White05} for mammals, \citep{Starostova09} for
cell size effects, and \citep{Hendriks07} for a review. At least
part of the controversy stems from the fact that the parameter $P_{0}$
is unlikely to be a true constant. Metabolic theory argues convincingly
that $P_{0}$ should be a function of the temperature \citep{Brown04}.
Both data and theory suggest that several other factors affect $P_{0}$\citep{Glazier10}.
Since the focus is on metabolism, the whole debate, in a more or less
implicit way, is framed by the assumption that the individuals live
in an 'ideal' environment, where 'ideal' means that all their needs
are fully satisfied.

In the last decade the ecological implications of body size has raised
a growing attention in the ecological literature and Kleiber-like
equations have been used in order to describe the rate of resource
intake as a function of body size, at least as a first approximation
\citep{Peters83,Brown04}. Yet, when applied to resource intake rate,
equation (\ref{eq:kleiber}) overlooks the role of individual adjustments
to scarcity of resources in a natural environment. The relationship
between intake of resources and their availability is addressed by
Holling's functional responses \citep{Holling59a,Holling59b}; furthermore,
the individual perception of resource availability has a dependency
on body size \citep{Ritchie98,Haskell02,Basset07}.%
{}

In a homogeneous environment, characterized by a given amount $R_{a}$
of available resources, the abundance of resources $R_{p}$ perceived
by any individual organism is a function of its size: larger organisms
may feel a sense of scarcity, while smaller ones still have a subjective
perception of abundance. In this context, 'homogeneous environment'
is a place where the spatial location of resources is unimportant,
and a single number ($R_{a}$) is sufficient to characterize the resources
available in that environment. We shall assume that the functional
link between perceived resources and mass is a power law: \begin{equation}
R_{p}=cR_{a}\left(\frac{M}{M_{0}}\right)^{-x}\label{eq:Rp}\end{equation}
where the value of the constant $x$ likely ranges from $1/4$ to
$3/4$ \citep{Basset07}, and $c$ is the normalization factor required
to match the available level of resources $R_{a}$ to the perceived
level of resources of an individual having exactly the baseline mass
$M_{0}$. 

How does an individual react to relative degrees of resource scarcity?
The traditional approach is to use the Holling's functional response
models, where a prescribed function $I$ links resource intake to
resource availability, even though several other reasonable choices
are possible for $I$ (i.e. \citealp{May72}). According to the cited
recent evidence suggesting that individual behaviour is primarily
affected by perceived resource availability rather than by absolute
availability, Holling's responses can be formulated as\begin{equation}
I=\frac{R_{p}^{\gamma}}{b^{\gamma}+R_{p}^{\gamma}}\label{eq:Holling}\end{equation}
where $\gamma\ge1$ ($\gamma=1$ is Holling type II, $\gamma=2$ is
Holling type III). Here the intake function is a non-dimensional quantity,
that ranges between zero and one. The independent variable is the
perceived resource level $R_{p}$, and the half-saturation coefficient
$b$ does not depend on the body size. The same intake function may
be expressed in a mathematically equivalent way by using $R_{a}$
as the independent variable, which is operationally more convenient
since $R_{a}$ is much easier to quantify experimentally than $R_{p}$.
In fact substituting (\ref{eq:Rp}) in (\ref{eq:Holling}) we obtain\begin{equation}
I=\frac{R_{a}^{\gamma}}{\left[b\left(M/M_{0}\right)^{x}c^{-1}\right]^{\gamma}+R_{a}^{\gamma}}.\label{eq:I(Ra)}\end{equation}
where the half-saturation coefficient is the mass-dependent function
$b_{a}(M)=b\left(M/M_{0}\right)^{x}c^{-1}$.

For less-than-ideal, or {}``natural'' conditions, the allometric
scaling law (\ref{eq:kleiber}), intended as a model of the resource
intake rate, needs to be corrected by the mass-dependent intake function
$I$ as as follows\begin{equation}
\frac{dQ}{dt}=P_{0}\left(\frac{M}{M_{0}}\right)^{\alpha}I(M)\label{eq:kleiber_I}\end{equation}
where $I$ is referred as $I(M)$ in order to make explicit the dependency
on body size and $Q$, from now on, is the mass of ingested resources.

\begin{figure}
\centering{}\includegraphics[width=0.99\columnwidth]{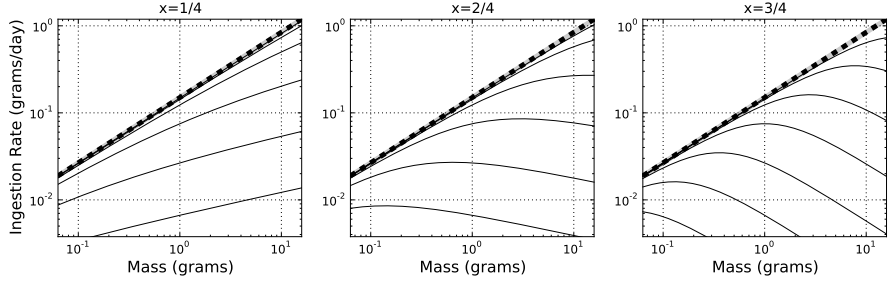}\caption{\label{fig:k-b}Plot of (\ref{eq:kleiber-basset}) for $\hat{R}_{a}=10^{-1},10^{-2/3},10^{-1/3},\cdots,10^{1}$
(thin black lines, $\hat{R}_{a}$ increasing upward). The thick dashed
line is the asymptotic law (\ref{eq:Abundance}); the exponent of
the perceived resources, from the left to the right panel is $x=1/4,\,2/4,\,3/4$.}

\end{figure}

At this point, we rewrite (\ref{eq:kleiber_I}) using (\ref{eq:I(Ra)})
to describe the quantitative relationship of intake rate with mass
and available resources:\begin{equation}
\frac{dQ}{dt}=P_{0}\frac{\hat{M}^{(\alpha-\gamma x)}}{\hat{R}_{a}^{-\gamma}+\hat{M}^{-\gamma x}}\label{eq:kleiber-basset}\end{equation}
where for convenience we have defined $\hat{M}=M/M_{0}$ and $\hat{R}_{a}=cR_{a}/b$.
In equation (\ref{eq:kleiber-basset}) the intake rate depends both
on body size and on resource availability. For every small interval
of $\hat{M}$ values, the function can be approximated by a scaling
law, whose scaling exponent is lower than the one appearing in (\ref{eq:kleiber}),
and decreases with increasing $\hat{M}$. This new relationship has
the following two interesting limits, both expressed by scaling laws:\begin{eqnarray}
\hat{M}\to\,0\, & \Rightarrow & \left.\frac{dQ}{dt}\right|_{\mathrm{Abundant}}\sim P_{0}\hat{M}^{\alpha}\hphantom{^{(-\gamma x)}\hat{R}_{a}^{\gamma}}\label{eq:Abundance}\end{eqnarray}
\begin{eqnarray}
\hat{M}\to\infty & \Rightarrow & \left.\frac{dQ}{dt}\right|_{\mathrm{Scarce}}\sim P_{0}\hat{R}_{a}^{\gamma}\hat{M}^{(\alpha-\gamma x)}\label{eq:Scarcity}\end{eqnarray}
where the symbol {}``$\sim$'' is used with its formal mathematical
meaning of {}``asymptotic to'' and the labels 'Abundant' and 'Scarce'
refer to size-dependent perceived resource availability. The upper
bound (\ref{eq:Abundance}) corresponds to the allometric law (\ref{eq:kleiber}):
for any assigned value of the resources $R_{a}$, there is a range
of small enough body sizes that perceives an unlimited 'ideal' environment.
At the opposite end of the spectrum of sizes, for any assigned value
of the resources $R_{a}$, the individuals are so large to be limited
by the perceived resource abundance so that (\ref{eq:Scarcity}) applies.
This is a scaling law in which the intake rate is proportional to
the $(\alpha-\gamma x)$-power of the body size, and also depends
on the overall amount of available resources (Figure \ref{fig:k-b}).
For large values of $\gamma x$, the scarcity regime may even yield
a power law in which the intake rate \emph{decreases} with the mass. 

In Figure \ref{fig:k-b} the quantity $Q$ represents the ingested
mass, $dQ/dt$ the ingested mass per unit of time ($P_{0}$ is also
mass per unit of time), the model parameter $M_{0}$ is posed equal
to one gram, and $P_{0}$ is approximated to 0.15 g/day according
to the experimental evidence \citep{Peters83,Basset92}. The exponent
$\gamma$ is set equal to 2, i.e. Holling type III functional response,
and the different curves plotted in each graph correspond to a different
value of $\hat{R}_{a}$. To mirror the uncertainty in the value of
$x$, we use the three $x$ values $x=1/4;$ $x=2/4;$ $x=3/4$. Therefore,
 the black lines represent the ingestion rates of individuals of varying
masses in an hypothetical experiment in which the level of available
resources is kept constant by replenishing the resources as the organism
consumes them (for example with a chemostat-like set-up). 

For high values of $x$ the relationship between resource ingestion
and body size shows a marked deviation from a power-law behaviour.
Of course, one should not expect to be able to observe the full asymptotic
regime (\ref{eq:Scarcity}), because it implicitly assumes that an
organism can sustain arbitrarily low intake rates, which is impossible.
When the available resources are too low, an organism must leave the
patch, migrate away from the resource-depleted region, or die. 

We also observe that in the present formulation, the relevant measure
of available resources is $\hat{R}_{a}$, not $R_{a}$. Since it is
$\hat{R}_{a}=cR_{a}/b$, it follows that among species with individuals
of comparable size, those characterized by a low value of $b$ remain
closer to the allometric law (\ref{eq:kleiber}) than those having
a larger value of $b$. In this sense, organisms with low $b$ cope
better with scarcity of resources. A similar argument holds for $c$.

\section{Case Studies on the Model Assumptions}

The model presented in the previous section is embodied by equation
(\ref{eq:kleiber-basset}) which depends on the validity of equation
(\ref{eq:Rp}). In this section we argue that these equations generalize
and extend concepts and ideas that have already been expressed in
the literature. We also address the consistency of the key model assumptions
with natural conditions using laboratory experiments and a metadata
analysis of case studies.

Equation (\ref{eq:Rp}) describes mathematically the decrease of perceived
resource availability as the individual body-size increases, given
a fixed level of absolute resources. Such an inverse relationship
was already implicitly incorporated in different models dealing with
individual patch choice dynamics \citep{Ritchie98}, home range size
\citep{Haskell02}, coexistence relationships \citep{Basset07}, but
it was not explicitly modelled, so far. An individual-based perception
of resources dates back to the \citet{MacArthur64} environmental
grain concept and it is not limited to a body-size dependency. Actually
body size accounts only for part of the possible deviation; resource
distribution \citep{Haskell02}, resource defence mechanisms \citep{Abrams96},
individual consumer niche breadth \citep{Rossi85}, searching and
pursuit ability \citep{Krebs97}, risk adverse behavioural strategy
\citep{Werner1981} represent other sources of deviations of perceived
resource availability. However, body size represents a systematic
source of deviation of perceived resource availability, whose influence
can be modeled as a scaling law. In our model, body-size independent
and body-size dependent forcing factors can be described by allowing
for variations respectively in the coefficient $c$ and $x$ of the
scaling law (\ref{eq:Rp}). Accounting for the allometric variation
of perceived resource availability increases the realism of resource
availability assessment, even though it does not completely resolve
all other biases listed above.

Perceived resource availability is a more realistic but less easily
tractable measure of resource availability than overall biomass or
units of potential resources per unit of space. Estimates of perceived
resource availability can be indirectly derived from the analysis
of the patterns of variation of quantifiable patch use components
with individual body size. Here, we consider the half-saturation coefficient
of Holling\textquoteright{}s functional responses and giving-up-density
from resource patches (GUD) as quantifiable proxies for the resource
availability perceived at the individual level with increasing individual
body size. 

\begin{figure}
\centering{}\includegraphics[width=0.8\columnwidth]{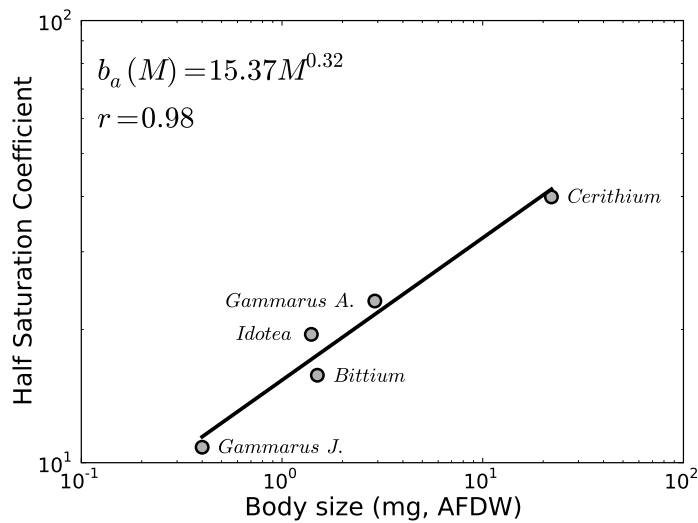}\caption{\label{fig:dati_basset}Allometric variation of the half saturation
coefficient $b_{a}$ of Holling Type III functional response with
body size in a guild of benthic detritivores. Data are from laboratory
experiments carried out using $^{32}\mathrm{P}$ labelled resources.
For every taxon or size class, food intake rate was assessed as the
$^{32}\mathrm{P}$ body burden in laboratory experiments where resource
availability ranged from 2 to 256 units of resources. Each unit was
represented by a single alder leaf disc, fully conditioned by micro-organisms,
which was previously labelled with $^{32}\mathrm{P}$ orthophosphate.
Techniques for $^{32}\mathrm{P}$ labelling of alder leaf discs are
described in Basset 1993. According to the model (\ref{eq:I(Ra)})
the size-dependent half saturation coefficient is $b_{a}(M)=b\left(M/M_{0}\right)^{x}c^{-1}$.}

\end{figure}

The half-saturation coefficient expresses the resource density at
which the individual resource intake rate is reduced to 50\% of the
maximum rate; the higher is the resource density at which individual
intake rate is reduced by 50\%, the lower is the perceived resource
density. In our conceptual framework the scaling of the perceived
resources with body size implies that the half-saturation coefficient
of Holling's functional responses must also scale as a power law of
the body size (equations \ref{eq:Holling} and \ref{eq:I(Ra)} combined
to give equation \ref{eq:kleiber_I}). In a case study we have run
experiments under laboratory conditions with benthic macro-invertebrates
in order to evaluate the behaviour of the functional response parameters
with species body size. According to the Holling Type III functional
response utilised in our model, the half saturation coefficient was
actually found to scale positively with individual body size with
a scaling factor not far from \textonequarter{} (i.e., $x=0.32$,
Figure \ref{fig:dati_basset} %
). This direct relationship was independent of the functional response
equation used; depending on the type of Holling\textquoteright{}s
functional response used for the fit, the half-saturation coefficient
scaled with species body-size with a positive exponent in a range
between \textonequarter{} and \textonehalf{}. A positive scaling factor
of the half-saturation coefficient with individual cell size was also
found on phytoplankton, where available data ($x=0.17$, \citealp{Valiela84})
showed a scaling factor close to \textonequarter{} . A range of values
between \textonequarter{} and \textthreequarters{} was also recently
used to investigate the influence of size dependent space use consumer
behaviour on species interaction and coexistence within competitive
guilds \citep{Basset07}. 

The occurrence of an inverse scaling of perceived resource availability
with body size is also derived by the analysis of published data on
the patch departure behaviour, using the resource giving up density
as a proxy for perceived resource availability: i.e., everything else
being equal, higher GUDs indicate lower perceived availabilities.
Actually, GUD data on the seed-eating rodent guilds showed higher
GUDs for higher individual body masses and metabolism (\citealp{Brown88,Brown94,Kotler02};
but see also: \citealp{Kotler93} for opposite evidence). Predation
risks (manipulation of predators: \citealp{Mohr03}; availability
of refugia and intensity of light: \citealp{Brownetal88}) were other
factors found to affect the GUD of seed-eating rodents. Therefore,
the assumption underlying equation (\ref{eq:Rp}) is supported by
a specific laboratory case study and by literature data from guilds
including invertebrate and vertebrate species. 

Equation (\ref{eq:kleiber-basset}) is a new model for the relationship
between intake rate and individual body-size, which admits a dependency
on resource availability. It is not a power law, but, for limited
ranges of body sizes, it can be approximated by power laws having
an exponent which decreases with increasing mass, and which tends
to a Kleiber-like law for un-limiting resource availability. So, equation
(\ref{eq:kleiber-basset}) summarizes and extends existing evidence
on deviations of the allometric exponent of the intake-rate versus
body-size relationship under limiting conditions. 

\begin{figure}
\centering{}\includegraphics[width=0.99\columnwidth]{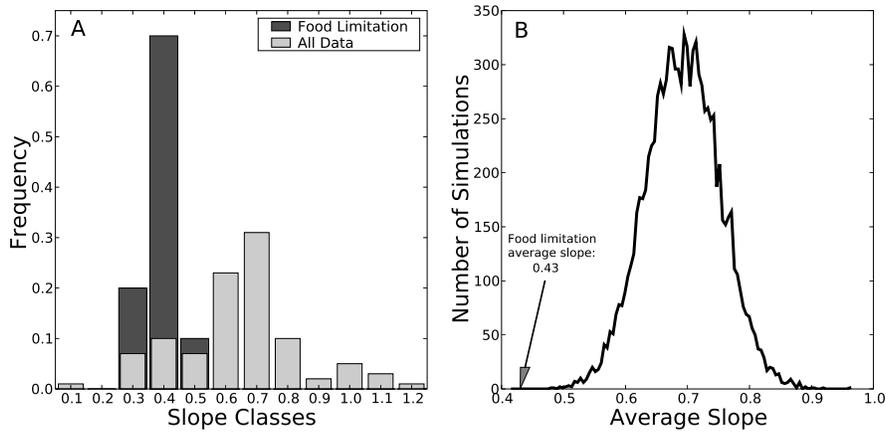}\caption{\label{fig:metadata}Meta-analysis of published data on the allometric
relationships between ingestion rates and body size in consumers including
aquatic and terrestrial groups, invertebrates and vertebrates. Data
are from 100 allometric regressions reported in the 51 published papers
listed in appendix \ref{sec:Appendix}. Ten regressions explicitly
refer to limiting conditions. Panel A: the frequency distribution
of slope values $\alpha$ of the allometric regressions are plotted,
comparing food limitation conditions with the overall data set. Panel
B: statistical comparison of the observed average slope in the food
limitation conditions with the results of 9999 Monte Carlo simulation
of 10 cases randomly selected from the overall data-set.}

\end{figure}

Literature data show a large variability of the scaling exponent of
resource intake-rate versus individual body-size, spanning a range
between $\alpha=0.1$ and $\alpha=1.2$, as shown in Figure \ref{fig:metadata}A
. The data refer to 100 experimental cases covering terrestrial and
aquatic ecosystems, invertebrate and vertebrate guilds. If the intake
rate always mirrored Kleiber's allometric law, such a large range
of experimental conditions would not be relevant for the observed
variability, since this law is supposed to cover a wide range of sizes
and taxonomic variability, and it is proposed as an universal law.
In our model for resource intake rates, an allometric law is just
an upper threshold for \textquoteleft{}ideal\textquoteright{} unlimited
conditions. The subsample of ten experimental cases specifically referring
to limiting conditions shows an average exponent $\alpha=0.43$, much
lower than the average exponent of the complete data set. A Montecarlo
simulation carried out by extracting 9999 randomly chosen subsamples
of ten elements shows, with an extremely high confidence level, that
the difference of the averages is statistically significant, and it
is not due to stochastic fluctuations (Figure \ref{fig:metadata}B).
Therefore, our model appears to be consistent with the evidence available
in the literature.

The patterns shown in Figure \ref{fig:k-b} go beyond the adaptation
of the individual ingestion rate to a low perceived level of resource
density at the local, patch, scale. For every fixed absolute resource
density occurring in a certain determined ecosystem, the corresponding
curve in the graph of Figure \ref{fig:k-b} describes how individual
fitness is affected by individual size. In fact, by increasing size
while the absolute resource density remains constant, individuals
need to cope with an increasingly strong resource limitation, by adopting
a suitable patch selection and departure behaviour, or by a home range
expansion \citep{Carbone02}, or by restricting the range of colonised
ecosystems according to the ecosystem size and overall productivity
\citep{Marquet05}; this hierarchy of implications of the size dependency
of resource limitation is then consistent with the observed patterns
of increasing extinction risk with increasing body size in vertebrates
\citep{Clauset09}.

\section{Conclusions}

The model described in this paper synthesizes in a single equation
three main factors of the process of animal resource intake: (a) the
dependence of individual metabolic requirements on individual body
size; (b) the dependence of individual resource intake rates on resource
availability; and, finally (c) the dependence of resource availability
perceived at the individual level on individual body size. It extends
the fields of application of both Holling and Kleiber equations. As
regards the former, the substitution of the term \textquoteleft{}resource
availability\textquoteright{} with the body-size dependent term \textquoteleft{}perceived
availability\textquoteright{} extends the application of Holling\textquoteright{}s
functional responses to the analysis of intra-guild competitive ability
and coexistence relationship. As regards the latter, the introduction
of the normalized intake function $I$ in the Kleiber equation extends
it to limiting conditions, incorporating conceptually a size dependency
in the allometric scaling coefficient. Finally, by unifying the components
of resource intake-rate variation due to individual body-size and
resource availability, the model contributes to the integration of
metabolic theory and resource perception with the dynamics of resource
availability.

\section{Acknowledgements. }

The research was funded by PRIN 2005 and FP7-WISER project grants
to A. Basset. We thank L. Rossi and D. Mouillot for their comments
on an earlier draft of this paper.

\section{Online Appendix\label{sec:Appendix}}

Literature data sources for Figure \ref{fig:metadata}. The key information
on the allometric scaling of individual resource intake-rate with
individual body-size are reported by listing the taxonomic group considered
({}``Group''), the slope value of the allometric relationship ({}``Exponent
$\alpha$''), the declared occurrence of resource limitation to the
consumers ({}``Limited'') and the reference source ({}``Reference'').

\noindent \begin{longtable}{>{\raggedright}p{0.3\columnwidth}lc>{\raggedright}p{0.5\columnwidth}}
Group  & Exponent $\alpha$  & Limited  & Reference\tabularnewline
\emph{\small Crassostea gigas}{\small{} } & {\small 0.19 } &  & {\small Bougrier et al., Aquaculture 134 (1995)}\tabularnewline
\emph{\small Crassostea gigas}{\small{} } & {\small 0.311 } &  & {\small Bougrier et al., Aquaculture 134 (1995)}\tabularnewline
\emph{\small Crassostea gigas}{\small{} } & {\small 0.312 } &  & {\small Bougrier et al., Aquaculture 134 (1995)}\tabularnewline
\emph{\small Crassostea gigas}{\small{} } & {\small 0.326 } &  & {\small Bougrier et al., Aquaculture 134 (1995)}\tabularnewline
\emph{\small Crassostea gigas}{\small{} } & {\small 0.348 } &  & {\small Bougrier et al., Aquaculture 134 (1995)}\tabularnewline
{\small Grazer ruminants } & {\small 0.36 } & {\small x} & {\small Illius\&Gordon, Journal of Animal Ecology 56 (1987)}\tabularnewline
{\small Grazer mammals } & {\small 0.36 } & {\small x} & {\small Clutton-Brock\&Harvey, Special Publication of the American
Society of Mammologists 7 (1983)}\tabularnewline
\emph{\small Crassostea gigas } & {\small 0.364 } &  & {\small Bougrier et al., Aquaculture 134 (1995)}\tabularnewline
\emph{\small Mytilus edulis } & {\small 0.408 } & {\small x} & {\small Thompson, Marine Biology 79 (1984)}\tabularnewline
{\small Rotifers } & {\small 0.417 } & {\small x} & {\small Stemberger\&Gilbert, Ecology 68 (1987)}\tabularnewline
{\small Daphniids } & {\small 0.42 } & {\small x} & {\small Jeyasingh, Ecology Letters 10 (2007)}\tabularnewline
{\small Subantartic copepods } & {\small 0.42 } & {\small x} & {\small Atkinson, Marine Ecology Progress Series 130 (1996)}\tabularnewline
\emph{\small Crassostea gigas } & {\small 0.422 } &  & {\small Bougrier et al., Aquaculture 134 (1995)}\tabularnewline
{\small Subantartic copepods } & {\small 0.43 } & {\small x} & {\small Atkinson, Marine Ecology Progress Series 130 (1996)}\tabularnewline
\emph{\small Ursus arctos horribilis } & {\small 0.44 } &  & {\small Rode et al., Oecologia 128 (2001)}\tabularnewline
{\small Pinnipeds Adult } & {\small 0.44 } &  & {\small Innes et al., Journal of Animal Ecology 1987}\tabularnewline
{\small Grazers mammals } & {\small 0.48 } & {\small x} & {\small Conradt et al., Animal Behaviour 59 (2000)}\tabularnewline
{\small Subantartic copepods } & {\small 0.49 } & {\small x} & {\small Atkinson, Marine Ecology Progress Series 130 (1996)}\tabularnewline
{\small Subantartic copepods } & {\small 0.51 } & {\small x} & {\small Atkinson, Marine Ecology Progress Series 130 (1996)}\tabularnewline
\emph{\small Crassostea gigas } & {\small 0.535 } &  & {\small Bougrier et al., Aquaculture 134 (1995)}\tabularnewline
\emph{\small Crassostea gigas } & {\small 0.539 } &  & {\small Bougrier et al., Aquaculture 134 (1995)}\tabularnewline
\emph{\small Ursus arctos horribilis } & {\small 0.57 } &  & {\small Rode et al., Oecologia 128 (2001)}\tabularnewline
{\small Terrestrial carnivora adult } & {\small 0.58 } &  & {\small Innes et al., Journal of Animal Ecology 1987}\tabularnewline
{\small Mustelidae adult } & {\small 0.58 } &  & {\small Innes et al., Journal of Animal Ecology 1987}\tabularnewline
\emph{\small Crassostea gigas } & {\small 0.585 } &  & {\small Bougrier et al., Aquaculture 134 (1995)}\tabularnewline
{\small Arid zone marsupials } & {\small 0.601 } &  & {\small Nagy\&Bradshaw, Journal of Mammalogy 81 (2000)}\tabularnewline
\emph{\small Chlamys nobilis } & {\small 0.601 } &  & {\small Pan\&Wang, Marine Ecology Progress Series 365 (2008)}\tabularnewline
\emph{\small Dreissena polymorpha } & {\small 0.61 } &  & {\small Schneider et al., Oecologia 117 (1998)}\tabularnewline
\emph{\small Chlamys farreri } & {\small 0.62 } &  & {\small Bacher et al., Aquating Living Resources 16 (2003) }\tabularnewline
\emph{\small Alces alces calves } & {\small 0.62 } &  & {\small Andersen\&Saether, Ecology 73 (1992) }\tabularnewline
{\small Copepods } & {\small 0.623 } &  & {\small Ikeda, Journal of Experimental Marine Ecology and Biology
29 (1977)}\tabularnewline
{\small Raptorial birds } & {\small 0.63 } &  & {\small Calder\&King, Avian Biology IV (1974) }\tabularnewline
{\small Daphniids } & {\small 0.63 } &  & {\small Jeyasingh, Ecology Letters 10 (2007)}\tabularnewline
{\small Raptorial birds } & {\small 0.63 } &  & {\small Schoener, The American Naturalist 49 (1968)}\tabularnewline
{\small Pinnipeds adult and terrestrial carnivora } & {\small 0.63 } &  & {\small Innes et al., Journal of Animal Ecology 1987}\tabularnewline
\emph{\small Crassostea gigas } & {\small 0.64 } &  & {\small Bougrier et al., Aquaculture 134 (1995)}\tabularnewline
{\small Herbivorous caecum fermenters } & {\small 0.64 } &  & {\small Clauss et al., Comparative Biochemistry and Physiology (2007)}\tabularnewline
\emph{\small Crassostea gigas } & {\small 0.663 } &  & {\small Bougrier et al., Aquaculture 134 (1995)}\tabularnewline
\emph{\small Crassostea gigas } & {\small 0.67 } &  & {\small Bougrier et al., Aquaculture 134 (1995)}\tabularnewline
{\small Delphinoidea } & {\small 0.67 } &  & {\small Innes et al., Marine Mammal Sciences 2 (1986)}\tabularnewline
{\small Forest floor arthropods } & {\small 0.68 } &  & {\small Reichle, Ecology 49 (1968)}\tabularnewline
{\small Mammals } & {\small 0.68 } &  & {\small Harestad\&Bunnel, Ecology 60 (1979)}\tabularnewline
{\small Pleuronectes platessa } & {\small 0.68 } &  & {\small Van der Veer et al., Journal of Sea Research in press (2009)}\tabularnewline
{\small Sea Ducks } & {\small 0.69 } &  & {\small Goudie\&Ankney, Ecology 67 (1986)}\tabularnewline
{\small Carnivorous homeotherms } & {\small 0.692 } &  & {\small Farlow, Ecology 57 (1976)}\tabularnewline
{\small Invertebrates } & {\small 0.694 } &  & {\small Capriulo, Marine Biology 71 (1982)}\tabularnewline
\emph{\small Crassostea gigas } & {\small 0.695 } &  & {\small Bougrier et al., Aquaculture 134 (1995)}\tabularnewline
{\small Carnivorous mammals } & {\small 0.697 } &  & {\small Farlow, Ecology 57 (1976)}\tabularnewline
\emph{\small Styela plicata } & {\small 0.7 } &  & {\small Fisher, Marine Biology 41 (1977)}\tabularnewline
{\small Passerine birds } & {\small 0.7 } &  & {\small Lindstrom\&Kvist, Proceedings Biological Sciences 261 (1995)}\tabularnewline
{\small Herbivorous mammals } & {\small 0.7 } &  & {\small Shipley et al., The American Naturalist 143 (1994)}\tabularnewline
{\small Capitella sp. } & {\small 0.701 } &  & {\small Forbes\&Lopez, Biological Bulletin 172 (1987) }\tabularnewline
{\small Homeotherms } & {\small 0.703 } &  & {\small Farlow, Ecology 57 (1976)}\tabularnewline
{\small Marine calanoid copepods } & {\small 0.703 } &  & {\small Saiz\&Calbet, Limnology and Oceanography 52 (2007)}\tabularnewline
\emph{\small Crassostea gigas } & {\small 0.707 } &  & {\small Bougrier et al., Aquaculture 134 (1995)}\tabularnewline
{\small Phocidae adults } & {\small 0.71 } &  & {\small Innes et al., Journal of Animal Ecology 1987}\tabularnewline
{\small Herbivorous homeoterms } & {\small 0.716 } &  & {\small Farlow, Ecology 57 (1976)}\tabularnewline
{\small Mammals and birds } & {\small 0.72 } &  & {\small Kirkwood, Comparative Biochemistry and Physiology 75 (1983)}\tabularnewline
{\small Phocidae juveniles } & {\small 0.72 } &  & {\small Innes et al., Journal of Animal Ecology 1987}\tabularnewline
{\small Phocidae adults } & {\small 0.72 } &  & {\small Innes et al., Journal of Animal Ecology 1987}\tabularnewline
{\small Herbivorous mammals } & {\small 0.728 } &  & {\small Farlow, Ecology 57 (1976)}\tabularnewline
{\small Terrestrial mammals } & {\small 0.73 } &  & {\small Nagy et al., Annual Review of Nutrition 19 (1999)}\tabularnewline
\emph{\small Alces alces}{\small{} adults } & {\small 0.73 } &  & {\small Andersen\&Saether, Ecology 73 (1992) }\tabularnewline
{\small Deposit-feeders } & {\small 0.74 } &  & {\small Cammen, Estuaries and Coasts 3 (1980a)}\tabularnewline
{\small Phocidae juveniles } & {\small 0.74 } &  & {\small Innes et al., Journal of Animal Ecology 1987}\tabularnewline
{\small Benthic detritivores } & {\small 0.742 } &  & {\small Cammen, Oecologia 44 (1980b)}\tabularnewline
{\small Whales } & {\small 0.75 } &  & {\small Hinga, Deep Sea Research 26 A (1979)}\tabularnewline
{\small Marine amphipods } & {\small 0.75 } &  & {\small Dagg, Internationale Revue der Gesamten Hydrobiologie 61 (1976)}\tabularnewline
{\small Zoo mammals } & {\small 0.75 } &  & {\small Evans\&Miller, Proceedings of the Nutrition Society 27 (1968)}\tabularnewline
{\small Cattles } & {\small 0.75 } &  & {\small Murray, The Journal of Animal Ecology 60 (1991)}\tabularnewline
{\small Herbivorous non-ruminant foregut fermenters } & {\small 0.76 } &  & {\small Clauss et al., Comparative Biochemistry and Physiology (2007)}\tabularnewline
{\small Herbivorous mammals } & {\small 0.76 } &  & {\small Clauss et al., Comparative Biochemistry and Physiology (2007) }\tabularnewline
{\small Herbivorous mammals } & {\small 0.77 } &  & {\small Clauss et al., Comparative Biochemistry and Physiology (2007) }\tabularnewline
{\small Zooplankton } & {\small 0.77 } &  & {\small Hansen et al., Limnology and Oceanography 42 (1997)}\tabularnewline
{\small Ungulates } & {\small 0.77 } &  & {\small Clauss et al., Comparative Biochemistry and Physiology (2007) }\tabularnewline
{\small Carnivores } & {\small 0.77 } &  & {\small Carbone et al., PLoS Biology 5 2007}\tabularnewline
{\small Geese } & {\small 0.78 } &  & {\small Durant et al., Journal of Animal Ecology 72 (2003)}\tabularnewline
{\small Herbivorous colon fermenters } & {\small 0.79 } &  & {\small Clauss et al., Comparative Biochemistry and Physiology (2007) }\tabularnewline
{\small Benthic detritivores } & {\small 0.79 } &  & {\small Cammen, Oecologia 44 (1980b)}\tabularnewline
{\small Crustacea } & {\small 0.8 } &  & {\small Conover, Marine Ecology IV (1978)}\tabularnewline
{\small Carnivorous poikilotherms } & {\small 0.82 } &  & {\small Farlow, Ecology 57 (1976)}\tabularnewline
{\small Periphyton } & {\small 0.83 } &  & {\small Cattaneo\&Mosseau, Oecologia 103 (1995)}\tabularnewline
{\small Ciliates } & {\small 0.84 } &  & {\small Fenchel, Microbial Ecology 6 (1980)}\tabularnewline
{\small Fishes } & {\small 0.841 } &  & {\small McCann, Ecology 79 (1998)}\tabularnewline
{\small Larval Fish } & {\small 0.843 } &  & {\small MacKenzie et al., Marine Ecology Progress Series 67 (1990)}\tabularnewline
\emph{\small Ursus americana } & {\small 0.86 } &  & {\small Welch et al., Ecology 78 (1997)}\tabularnewline
{\small Non-mustelid carnivora adult } & {\small 0.87 } &  & {\small Innes et al., Journal of Animal Ecology 1987}\tabularnewline
{\small Phocidae adults } & {\small 0.87 } &  & {\small Innes et al., Journal of Animal Ecology 1987}\tabularnewline
{\small Non-mustelid carnivora adult } & {\small 0.89 } &  & {\small Innes et al., Journal of Animal Ecology 1987}\tabularnewline
{\small Larval Fish } & {\small 0.99 } &  & {\small MacKenzie et al., Marine Ecology Progress Series 67 (1990)}\tabularnewline
\emph{\small Daphnia ambigua } & {\small 0.991 } &  & {\small Lynch et al., Limnology and Oceanography 31 (1986)}\tabularnewline
{\small Pigs } & {\small 1 } &  & {\small Wellock et al., Journal of Animal Science 81 (2003)}\tabularnewline
\emph{\small Odocoileus hemionus } & {\small 1 } &  & {\small Hobbs, Wildlife Monographs 101 (1989)}\tabularnewline
\emph{\small Daphnia parvula } & {\small 1.001 } &  & {\small Lynch et al., Limnology and Oceanography 31 (1986)}\tabularnewline
{\small Finches } & {\small 1.02 } &  & {\small Calder\&King, Avian Biology IV (1974)}\tabularnewline
\emph{\small Acartia tonsa } & {\small 1.08 } &  & {\small Berggreen, Marine Biology 99 (1988)}\tabularnewline
{\small Benthic detritivores } & {\small 1.115 } &  & {\small Cammen, Oecologia 44 (1980b)}\tabularnewline
{\small Larval Fish } & {\small 1.162 } &  & {\small MacKenzie et al., Marine Ecology Progress Series 67 (1990)}\tabularnewline
\emph{\small Daphnia pulex } & {\small 1.198 } &  & {\small Lynch et al., Limnology and Oceanography 31 (1986)}\tabularnewline
\emph{\small Daphnia galatea mendotae } & {\small 1.243 } &  & {\small Lynch et al., Limnology and Oceanography 31 (1986)}\tabularnewline
\end{longtable}

\bibliographystyle{apalike}
\bibliography{Kleiber_refs}

\end{document}